\documentclass[letterpaper,10pt]{article}
\usepackage[T1]{fontenc}
\usepackage{parskip}
\usepackage[font={small}]{caption}
\usepackage[hidelinks]{hyperref}
\usepackage[margin=2.5cm]{geometry}
\usepackage{times}
\usepackage{amsmath,amssymb,amsthm}
\usepackage[affil-it]{authblk}
\usepackage{listings}
\usepackage{graphicx}

\date{}

\usepackage[round]{natbib}

\usepackage{algorithm}
\usepackage{algpseudocode}
\usepackage{subcaption}
\usepackage{wrapfig}

\algrenewcommand\algorithmicindent{0.5em}

\newlength\figureheight
\newlength\figurewidth

\usepackage{xcolor}
\usepackage{amsmath}
\usepackage{todonotes}
\usepackage{amssymb,mathtools,bbm}
\newcommand{\nb}[1]{\textcolor{red}{#1}}

\newcommand{\Expb}[2]{\mathbb{E}_{#1}\left[{#2}\right]}
\newcommand{\Reals}{\mathbb{R}}
\newcommand{\Transp}{\mathsf{T}}
\newcommand{\ii}{{(i)}}
\newcommand{\ji}{{(j)}}

\newcommand{\iip}{{i}}
\newcommand{\jip}{{j}}
\newcommand{\Nip}{{N}}
\newcommand{\nx}{{n_x}}
\newcommand{\nuu}{{n_u}}
\newcommand{\ny}{{n_y}}
\newcommand{\anc}{a}
\newcommand{\N}{\mathcal{N}}
\newcommand{\IW}{\mathcal{IW}}
\newcommand{\MN}{\mathcal{MN}}
\newcommand{\MNIW}{\mathcal{MNIW}}
\newcommand{\phivec}{{\bar{\varphi}}}
\newcommand{\Prb}[1]{\mathbb{P}({#1})}
\newcommand{\pfw}{\omega}

\newcommand{\Inda}[1]{\mathbbm{1}_{#1}}
\newcommand{\tr}{\mathrm{tr}}
\newcommand{\tra}[1]{\mathrm{tr}\left\{{#1}\right\}}
\newcommand{\saemQ}{\mathbb{Q}}
\newcommand{\emQ}{\mathcal{Q}}

\renewcommand\mid{\,\vert\,}

\title{A flexible state-space model for learning\\nonlinear dynamical systems}
\author[]{Andreas Svensson\thanks{\url{andreas.svensson@it.uu.se}}~}
\author[]{Thomas B. Sch\"{o}n\thanks{\url{thomas.schon@it.uu.se}}}
\affil[]{Department of Information Technology, Uppsala University}

\begin{document}

\maketitle

\textbf{Please cite this version:}

A. Svensson and Sch\"on, T. B. (2017). A flexible state-space model for learning nonlinear dynamical systems.\\ \textit{Automatica, 80}, page 189--199.

\begin{center}
\begin{minipage}{\linewidth}
\begin{lstlisting}[breaklines,basicstyle=\small\ttfamily]
@article{SvenssonSchon2017,
author    = {Svensson, Andreas and Sch\"{o}n, Thomas B.},
title     = {A flexible state-space model for learning nonlinear dynamical systems},
journal   = {Automatica},
year      = {2017},
volume    = {80},
pages     = {189--199}
}
\end{lstlisting}
\end{minipage}
\end{center}

\vspace{5em}

\begin{abstract}
We consider a nonlinear state-space model with the state transition and observation functions expressed as basis function expansions. The coefficients in the basis function expansions are learned from data. Using a connection to Gaussian processes we also develop priors on the coefficients, for tuning the model flexibility and to prevent overfitting to data, akin to a Gaussian process state-space model. The priors can alternatively be seen as a regularization, and helps the model in generalizing the data without sacrificing the richness offered by the basis function expansion. To learn the coefficients and other unknown parameters efficiently, we tailor an algorithm using state-of-the-art sequential Monte Carlo methods, which comes with theoretical guarantees on the learning. Our approach indicates promising results when evaluated on a classical benchmark as well as real data.
\end{abstract}

\vspace{10em}

{
	
	\footnotesize 
	\textbf{This research is financially supported} by the Swedish Research Council
	via the project \emph{Probabilistic modeling of dynamical systems}
	(contract number: 621-2013-5524) and the Swedish Foundation for
	Strategic Research (SSF) via the project \emph{ASSEMBLE}. 
	
}

\clearpage

\section{Introduction}

Nonlinear system identification \citep{Ljung:1999,Ljung:2010,SZL+:1995} aims to learn nonlinear mathemati\-cal models from data generated by a dynamical system. We will tackle the problem of learning nonlinear state-space models with only weak assumptions on the nonlinear functions, and make use of the Bayesian framework \citep{Peterka:1981} to encode prior knowledge and assumptions to guide the otherwise too flexible model.

Consider the (time invariant) state-space model
\begin{subequations}\label{eq:ssm}
	\begin{align}
		x_{t+1} &= f(x_t,u_t) + v_t, &v_t\sim \N(0,Q),\label{eq:ssm:a}\\
		y_t &= g(x_t,u_t) + e_t, &e_t\sim \N(0,R).\label{eq:ssm:b}
	\end{align}
\end{subequations}
The variables are denoted as the state\footnote{$v_t$ and $e_t$ are iid with respect to $t$, and $x_t$ is thus Markov.} $x_t \in \Reals^{\nx}$, which is not observed explicitly, the input $u_t \in \Reals^{\nuu}$, and the output $y_t \in \Reals^{\ny}$. We will learn the state transition function $f: \Reals^\nx \times \Reals^\nuu \mapsto \Reals^\nx$ and the observation function $g: \Reals^\nx \times \Reals^\nuu \mapsto \Reals^\ny$ as well as $Q$ and $R$ from a set of training data of input-output signals $\{u_{1:T},y_{1:T}\}$.

\begin{wrapfigure}[31]{r}{.43\linewidth}
	\begin{center}
		\includegraphics[width=\linewidth]{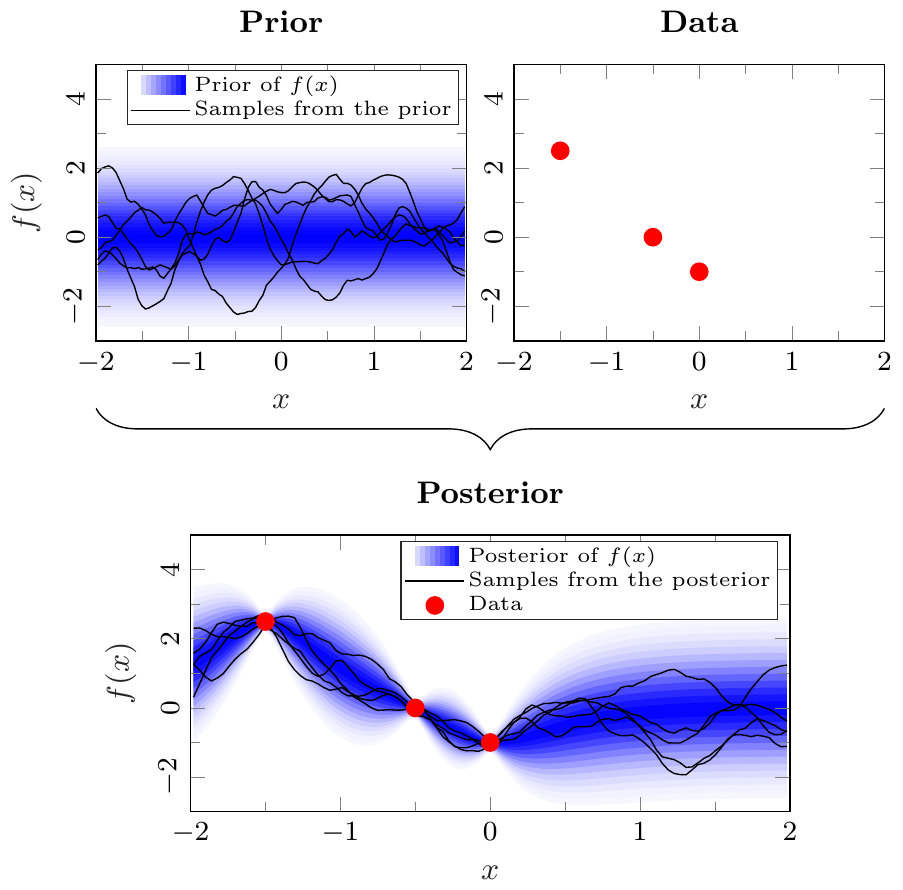}
		
		\caption{The Gaussian process as a modeling tool for an one-dimensional function $f: \Reals\mapsto\Reals$. The prior distribution (upper left plot) is represented by the shaded blue color (the more intense color, the higher density), as well as 5 samples drawn from it. By combining the prior and the data (upper right plot), the posterior (lower plot) is obtained. The posterior mean basically interpolates between the data points, and adheres to the prior in regions where the data is not providing any information. This is clearly a desirable property when it comes to generalizing from the training data---consider the thought experiment of using a 2nd order polynomial instead. Further, the posterior also provides a quantification of the uncertainty present, high in data-scarce regions and low where the data provides knowledge about $f(\cdot)$.}
		\label{fig:GP}
	\end{center}
\end{wrapfigure}

Consider a situation when a finite-dimensional linear, or other sparsely parameterized model, is too rigid to describe the behavior of interest, but only a limited data record is available so that any too flexible model would overfit (and be of no help in generalizing to events not exactly seen in the training data). In such a situation, a \textit{systematic way to encode prior assumptions and thereby tuning the flexibility of the model} can be useful. For this purpose, we will take inspiration from Gaussian processes (GPs, \citealt{RW:2006}) as a way to encode prior assumptions on $f(\cdot)$ and $g(\cdot)$. As illustrated by Figure~\ref{fig:GP}, the GP is a distribution over functions which gives a probabilistic model for inter- and extrapolating from observed data. GPs have successfully been used in system identification for, e.g., response estimation, nonlinear ARX models and GP state-space models \citep{PD:2010,Kocijan:2016,Frigola:2015}.

To parameterize $f(\cdot)$, we expand it using basis functions
\begin{equation}
	f(x) = \sum_{j=0}^{m} w^\ji \phi^\ji(x), \label{eq:bfe}
\end{equation}
and similarly for $g(\cdot)$. The set of basis functions is denoted by $\{\phi^\ji(\cdot)\}_{j=0}^m$, whose coefficients $\{w^\ji\}_{j=0}^m$ will be learned from data. By introducing certain \textit{priors} $p(w^\ji)$ \textit{on the basis function coefficients} the connection to GPs will be made, based on a Karhunen-Lo\`{e}ve expansion \citep{SS:2014}. We will thus be able to understand our model in terms of the well-established and intuitively appealing GP model, but still benefit from the computational advantages of the linear-in-parameter structure of~\eqref{eq:bfe}. Intuitively, the idea of the priors $p(w^\ji)$ is to keep $w^\ji$ `small unless data convinces otherwise', or equivalently, introduce a regularization of $w^\ji$.

To learn the model~\eqref{eq:ssm}, i.e., determine the basis function coefficients $w^\ji$, we tailor a learning algorithm using recent sequential Monte Carlo/particle filter methods \citep{SLD+:2015,KDS+:2015}. The learning algorithm infers the posterior distribution of the unknown parameters from data, and come with theoretical guarantees. We will pay extra attention to the problem of finding the maximum mode of the posterior, or equivalent, regularized maximum likelihood estimation.

Our contribution is the development of a flexible nonlinear state-space model with a tailored learning algorithm, which together constitutes a new nonlinear system identification tool. The model can either be understood as a GP state-space model (generalized allowing for discontinuities, Section~\ref{sec:model:sing}), or as a nonlinear state-space model with a regularized basis function expansion.

\section{Related work}

Important work using the GP in system identification includes impulse response estimation \citep{PD:2010,PCD:2011,COL:2012}, nonlinear ARX models \citep{KGB+:2005,BSW+:2016}, Bayesian learning of ODEs \citep{CGL:2009,WB:2014,MHH:2015} and the latent force model \citep{ALL:2013}. In the GP state-space model \citep{Frigola:2015} the transition function $f(\cdot)$ in a state-space model is learned with a GP prior, particularly relevant to this paper.
A conceptually interesting contribution to the GP state-space model was made by \citet{FLS+:2013}, using a Monte Carlo approach (similar to this paper) for learning. The practical use of \citet{FLS+:2013} is however very limited, due to its extreme computational burden. This calls for approximations, and a promising approach is presented by \citet{FCR:2014} (and somewhat generalized by \citet{MDD+:2015}), using inducing points and a variational inference scheme. Another competitive approach is \citet{SSS+:2016}, where we applied the GP approximation proposed by \citet{SS:2014} and used a Monte Carlo approach for learning (\citet{Frigola:2015} covers the variational learning using the same GP approximation). In this paper, we extend this work by considering basis function expansions in general (not necessarily with a GP interpretation), introduce an approach to model discontinuities in $f(\cdot)$, as well as including both a Bayesian and a maximum likelihood estimation approach to learning.

To the best of our knowledge, the first extensive paper on the use of a basis function expansion inside a state-space model was written by \citet{GR:1999}, who also wrote a longer unpublished version \citep{RG:2000}. The recent work by \citet{TDM:2015} resembles that of \citet{GR:1999} on the modeling side, as they both use basis functions with locally concentrated mass spread in the state space. On the learning side, \citet{GR:1999} use an expectation maximization (EM, \citealt{DLR:1977}) procedure with extended Kalman filtering, whilst \citet{TDM:2015} use particle Metropolis-Hastings \citep{ADH:2010}. There are basically three major differences between \citet{TDM:2015} and our work. We will (i) use another (related) learning method, particle Gibbs, allowing us to take advantage of the linear-in-parameter structure of the model to increase the efficiency. Further, we will (ii) mainly focus on a different set of basis functions (although our learning procedure will be applicable also to the model used by \citet{TDM:2015}), and -- perhaps most important -- (iii) we will pursue a systematic encoding of prior assumptions further than \citet{TDM:2015}, who instead assume $g(\cdot)$ to be known and use `standard sparsification criteria from kernel adaptive filtering' as a heuristic approach to regularization.

There are also connections to \citet{PLS+:2010}, who use a polynomial basis inside a state-space model. In contrast to our work, however, \citet{PLS+:2010} prevent the model from overfitting to the training data not by regularization, but by manually choosing a low enough polynomial order and terminating the learning procedure prematurely (early stopping). \citeauthor{PLS+:2010} are, in contrast to us, focused on the frequency properties of the model and rely on optimization tools. An interesting contribution by \citeauthor{PLS+:2010} is to first use classical methods to find a linear model, which is then used to initialize the linear term in the polynomial expansion. We suggest to also use this idea, either to initialize the learning algorithm, or use the nonlinear model only to describe deviations from an initial linear state-space model.

Furthermore, there are also connections to our previous work \citep{SSS+:2015}, a short paper only outlining the idea of learning a regularized basis function expansion inside a state-space model. Compared to \citet{SSS+:2015}, this work contains several extensions and new results. Another recent work using a regularized basis function expansion for nonlinear system identification is that of \citet{DAG+:2015}, however not in the state-space model framework. \citet{DAG+:2015} use rank constrained optimization, resembling an $L^0$-regularization. To achieve a good performance with such a regularization, the system which generated the data has to be well described by only a few number of the basis functions being `active', i.e., have non-zero coefficients, which makes the choice of basis functions important and problem-dependent. The recent work by \citet{MD:2016} is also covering learning of a regularized basis function expansion, however for input-output type of models.

\section{Constructing the model}\label{sec:model}
We want the model, whose parameters will be learned from data, to be able to describe a broad class of nonlinear dynamical behaviors without overfitting to training data. To achieve this, important building blocks will be the basis function expansion~\eqref{eq:bfe} and a GP-inspired prior. The order $\nx$ of the state-space model~\eqref{eq:ssm} is assumed known or set by the user, and we have to learn the transition and observation functions  $f(\cdot)$ and $g(\cdot)$ from data, as well as the noise covariance matrices $Q$ and $R$. For brevity, we focus on $f(\cdot)$ and $Q$, but the reasoning extends analogously to $g(\cdot)$ and $R$.

\subsection{Basis function expansion}

The common approaches in the literature on black-box modeling of functions inside state-space models can broadly be divided into three groups: neural networks \citep{Bishop:2006,NL:1996,NRP+:2000}, basis function expansions \citep{SZL+:1995,GR:1999,PLS+:2010,TDM:2015} and GPs \citep{RW:2006,Frigola:2015}. We will make use of a basis function expansion inspired by the GP. There are several reasons for this: Firstly, a basis function expansion provides an expression which is linear in its parameters, leading to a computational advantage: neural networks do not exhibit this property, and the na\"{i}ve use of the nonparametric GP is computationally very expensive. Secondly, GPs and some choices of basis functions allow for a straightforward way of including prior assumptions on $f(\cdot)$ and help generalization from the training data, also in contrast to the neural network.

We write the combination of the state-space model~\eqref{eq:ssm} and the basis function expansion~\eqref{eq:bfe} as
\begin{subequations}\label{eq:ss2}
\begin{align}
x_{t+1} &= \underbrace{\begin{bmatrix} w_{1}^{(1)} & \cdots & w_{1}^{(m)} \\ \vdots & & \vdots \\ w_{\nx}^{(1)} & \cdots & w_{\nx}^{(m)}  \end{bmatrix}}_{A}
\underbrace{\begin{bmatrix} \phi^{(1)}_{\vphantom{1}}(x_t,u_t) \\ \vdots \\ \phi^{(m)}_{\vphantom{\nx}}(x_t,u_t) \end{bmatrix}}_{\phivec(x_t,u_t)}
+ v_t,\\
y_{t} &= \underbrace{\begin{bmatrix} w_{g,1}^{(1)} & \cdots & w_{g,1}^{(m)} \\ \vdots & & \vdots \\ w_{g,\ny}^{(1)} & \cdots & w_{g,\ny}^{(m)}  \end{bmatrix}}_{C}
\underbrace{\begin{bmatrix} \phi^{(1)}_{g\vphantom{1}}(x_t,u_t) \\ \vdots \\ \phi^{(m)}_{g\vphantom{\nx}}(x_t,u_t) \end{bmatrix}}_{\phivec_g(x_t,u_t)}
+ e_t.
\end{align}
\end{subequations}
There are several alternatives for the basis functions, e.g., polynomials \citep{PLS+:2010}, the Fourier basis \citep{SSS+:2015}, wavelets \citep{SZL+:1995}, Gaussian kernels \citep{GR:1999,TDM:2015} and piecewise constant functions.
For the one-dimensional case (e.g., $\nx=1$, $\nuu=0$) on the interval $[-L,L] \in \Reals$, we will choose the basis functions as
\begin{align}
	\phi^\ji(x) = \frac{1}{\sqrt{L}}\sin\left(\frac{\pi j(x+L)}{2L}\right). \label{eq:bf}
\end{align}
This choice, which is the eigenfunctions to the Laplace operator, enables a particularly convenient connection to the GP framework \citep{SS:2014} in the priors we will introduce in Section~\ref{sec:prior:GP}. This choice is, however, important only for the interpretability\footnote{Other choices of basis functions are also interpretable as GPs. The choice~\eqref{eq:bf} is, however, preferred since it is independent of the choice of which GP covariance function to use.} of the model. The learning algorithm will be applicable to any choice of basis functions.

\subsubsection{Higher state-space dimensions}
The generalization to models with a state space and input dimension such that $\nx+\nuu > 1$ offers no conceptual challenges, but potentially computational ones. The counterpart to the basis function~\eqref{eq:bf} for the space\\$[-L_1,L_1] \times \dots \times [-L_{\nx+\nuu},L_{\nx+\nuu}] \in \Reals^{\nx+\nuu}$ is 
\begin{equation}
\phi^{(j_1,\dots,j_{\nx+\nuu})}(x) = \prod_{k=1}^{\nx+\nuu} \frac{1}{\sqrt{L_k}}\sin\left(\frac{\pi j_k(x^k\!+\!L_k)}{2L_k}\right), \label{eq:mdbf}
\end{equation}
(where $x^k$ is the $k$th component of $x$), implying that the number of terms $m$ grows exponentially with ${\nx+\nuu}$. This problem is inherent in most choices of basis function expansions. For $\nx>1$, the problem of learning $f: \Reals^{\nx+\nuu} \mapsto$ $\Reals^{\nx}$ can be understood as learning $\nx$ number of functions $f_i: \Reals^{\nx+\nuu} \mapsto \Reals$, cf.~\eqref{eq:ss2}. 

There are some options available to overcome the exponential growth with $\nx+\nuu$, at the cost of a limited capability of the model. \textit{Alternative 1} is to assume $f(\cdot)$ to be `separable' between some dimensions, e.g., $f(x_{t}, u_{t}) = f^x(x_t) + f^u(u_t)$. If this assumption is made for all dimensions, the total number of parameters present grows quadratically (instead of exponentially) with $\nx+\nuu$. \textit{Alternative 2} is to use a radial basis function expansion \citep{SZL+:1995}, i.e., letting $f(\cdot)$ only be a function of some norm $\|\cdot\|$ of $(x_t,u_t)$, as $f(x_{t}, u_{t}) = f(\|(x_{t}, u_{t})\|)$. The radial basis functions give a total number of parameters growing linearly with $\nx+\nuu$. Both alternatives will indeed limit the space of functions possible to describe with the basis function expansion. However, as a pragmatic solution to the otherwise exponential growth in the number of parameters it might still be worth considering, depending on the particular problem at hand.

\subsubsection{Manual and data-driven truncation}\label{sec:truncation}

To implement the model in practice, the number of basis functions $m$ has to be fixed to a finite value, i.e., truncated. However, fixing $m$ also imposes a harsh restriction on which functions $f(\cdot)$ that can be described. Such a restriction can prevent overfitting to training data, an argument used by \citet{PLS+:2010} for using polynomials only up to 3rd order. We suggest, on the contrary, to use priors on $w^\ji$ to prevent overfitting, and we argue that the interpretation as a GP is a preferred way to tune the model flexibility, rather than manually and carefully tuning the truncation. We therefore suggest to choose $m$ as big as the computational resources allows, and let the prior and data decide which $w^\ji$ to be nonzero, a \emph{data-driven truncation}.

Related to this is the choice of $L$ in \eqref{eq:bf}: if $L$ is chosen too small, the state space becomes limited and thereby also limits the expressiveness of the model. On the other hand, if $L$ is too big, an unnecessarily large $m$ might also be needed, wasting computational power. To chose $L$ to have about the same size as the maximum of $u_t$ or $y_t$ seems to be a good guideline.

\subsection{Encoding prior assumptions---regularization}\label{sec:prior}

The basis function expansion~\eqref{eq:ss2} provides a very flexible model. A prior might therefore be needed to generalize from, instead of overfit to, training data. From a user perspective, the prior assumptions should ultimately be formulated in terms of the input-output behavior, such as gains, rise times, oscillations, equilibria, limit cycles, stability etc. As of today, tools for encoding such priors are (to the best of the authors' knowledge) not available. As a resort, we therefore use the GP state-space model approach, where we instead encode prior assumptions on $f(\cdot)$ as a GP. Formulating prior assumptions on $f(\cdot)$ is relevant in a model where the state space bears (partial) physical meaning, and it is natural to make assumptions whether the state $x_t$ is likely to rapidly change (non-smooth $f(\cdot)$), or state equilibria are known, etc. However, also the truly black-box case offers some interpretations: a very smooth $f(\cdot)$ corresponds to a locally close-to-linear model, and vice versa for a more curvy $f(\cdot)$, and a zero-mean low variance prior on $f(\cdot)$ will steer the model towards a bounded output (if $g(\cdot)$ is bounded).

To make a connection between the GP and the basis function expansion, a Karhunen-Lo\`{e}ve expansion is explored by \citet{SS:2014}. We use this to formulate Gaussian priors on the basis function expansion coefficients $w^\ji$, and learning of the model will amount to infer the 
posterior $p(w^\ji|y_{1:T}) \propto p(y_{1:T}|w^\ji)p(w^\ji)$, where $p(w^\ji)$ is the prior and $p(y_{1:T}|w^\ji)$ the likelihood. To use a prior $w^\ji \sim \N(0,\alpha^{-1})$ and inferring the maximum mode of the posterior can equivalently be interpreted as regularized maximum likelihood estimation
\begin{align}
\arg\min_{w^\ji} ~-\log p(y_{1:T}|w^\ji) + \alpha|w^\ji|^2.
\end{align}

\subsubsection{Smooth GP-priors for the functions}\label{sec:prior:GP}

The Gaussian process provides a framework for formulating prior assumptions on functions, resulting in a non-parametric approach for regression. In many situations the GP allows for an intuitive generalization of the training data, as illustrated by Figure~\ref{fig:GP}. We use the notation
\begin{equation}
f(x)\sim\mathcal{GP}(m(x),\kappa(x,x'))
\end{equation}
to denote a GP prior on $f(\cdot)$, where $m(x)$ is the mean function and $\kappa(x,x')$ the covariance function.
The work by \citet{SS:2014} provides an explicit link between basis function expansions and GPs based on the Karhunen-Lo\`{e}ve expansion, in the case of isotropic\footnote{Note, this concerns only $f(\cdot)$, which resides \textit{inside} the state-space model. This does \emph{not} restrict the input-output behavior, from $u(t)$ to $y(t)$, to have an isotropic covariance.} covariance functions, i.e., $\kappa(x,x') = $ $ \kappa(|x-x'|)$. In particular, if the basis functions are chosen as~\eqref{eq:bf}, then
\begin{subequations}
\begin{equation}
f(x)\sim\mathcal{GP}(0,\kappa(x,x'))\Leftrightarrow f(x) \approx \sum_{j=0}^{m} w^\ji \phi^\ji(x),\label{eq:gpbfe}
\end{equation}
with\footnote{The approximate equality in~\eqref{eq:gpbfe} is exact if $m\to\infty$ and $L\to\infty$, refer to \cite{SS:2014} for details.}
\begin{equation}
w^\ji \sim \N(0,S(\lambda^\ji)),\label{eq:GPpriorw}
\end{equation}
\end{subequations}
where $S$ is the spectral density of $\kappa$, and $\lambda^\ji$ is the eigenvalue of $\phi^\ji$. Thus, this gives a systematic guidance on how to choose basis functions and priors on $w^\ii$. In particular, the eigenvalues of the basis function~\eqref{eq:bf} are
\begin{align}
	\lambda^\ji = \left(\frac{\pi j}{2 L}\right)^2, ~\text{and}~
	\lambda^{(j_{1:\nx+\nuu})} = \sum_{k=1}^{\nx+\nuu} \left(\frac{\pi j_k}{2 L_k}\right)^2
\end{align}
for~\eqref{eq:mdbf}. Two common types of covariance functions are the exponentiated quadratic $\kappa_{\text{eq}}$ and Mat\'{e}rn $\kappa_{\text{M}}$ class \citep{RW:2006},
\begin{subequations}\label{eq:covfs}
	\begin{align}
		\kappa_{\text{eq}}(r) &= s_f\exp\left( -\tfrac{r^2}{2l^2} \right),\label{eq:eq}\\
		\kappa_{\text{M}}(r) &= s_f\tfrac{2^{1-\nu}}{\Gamma(\nu)}\left(\tfrac{\sqrt{2\nu}r}{l}\right)^\nu K_\nu\left(\tfrac{\sqrt{2\nu}r}{l}\right),
	\end{align}
\end{subequations}
where $r \triangleq x-x'$, $K_\nu$ is a modified Bessel function, and $\ell, s_f$ and $\nu$ are hyperparameters to be set by the user or to be marginalized out, see \citet{SSS+:2016} for details. Their spectral densities are
\begin{subequations}
	\begin{align}
		S_{\text{eq}}(s) &= s_f\sqrt{2\pi l^2}\exp\left(-\tfrac{\pi^2l^2s^2}{2}\right), \\
		S_{\text{M}}(s) &= s_f\tfrac{2\pi^{\frac{1}{2}}\Gamma(\nu+\tfrac{1}{2})(2\nu)^\nu}{\Gamma(\nu)l^{2\nu}}\left(\tfrac{2\nu}{l^2} + s^2\right)^{-(\nu+\frac{1}{2})}.
	\end{align}
\end{subequations}
Altogether, by choosing the priors for $w^\ji$ as~\eqref{eq:GPpriorw}, it is possible to approximately interpret $f(\cdot)$, parameterized by the basis function expansion~\eqref{eq:bfe}, as a GP. For most covariance functions, the spectral density $S(\lambda^\ji)$ tends towards $0$ when $\lambda^\ji \to\infty$, meaning that the prior for large $j$ tends towards a Dirac mass at $0$. Returning to the discussion on truncation (Section \ref{sec:truncation}), we realize that truncation of the basis function expansion with a reasonably large $m$ therefore has no major impact to the model, but the GP interpretation is still relevant.

As discussed, finding the posterior mode under a Gaussian prior is equivalent to $L^2$-regularized maximum likelihood estimation. There is no fundamental limitation prohibiting other priors, for example Laplacian (corresponding to $L^1$-regularization: \citealt{Tibshirani:1996}). We use the Gaussian prior because of the connection to a GP prior on $f(\cdot)$, and it will also allow for closed form expressions in the learning algorithm.

For book-keeping, we express the prior on $w^\ji$ as a Matrix normal ($\MN$, \citealt{Dawid:1981}) distribution over $A$. The $\MN$ distribution is parameterized by a mean matrix ${M} \in \Reals^{\nx \times m}$, a right covariance ${U} \in \Reals^{\nx \times \nx}$ and a left covariance ${V} \in \Reals^{m \times m}$. The $\MN$ distribution can be defined by the property that $A \sim \MN(M,U,V)$ if and only if $\textrm{vec}(A)\sim\N(\textrm{vec}(M),V\otimes U)$, where $\otimes$ is the Kronecker product. Its density can be written as
\begin{equation}
	\mathcal{MN}({A}\mid {M},{U},{V}) = \frac{\exp\left(-\frac{1}{2}\mathrm{tr}\left\{({A}-{M})^\Transp U^{-1}({A}-{M}){V}^{-1}\right\}\right)}{(2\pi)^{\nx m}|{V}|^{\nx/2}|{U}|^{m/2}}. \label{eq:MNpdf}
\end{equation}
By letting $M = 0$ and $V$ be a diagonal matrix with entries $S(\lambda^\ji)$, the priors~\eqref{eq:GPpriorw} are incorporated into this parametrization. We will let $U=Q$ for conjugacy properties, to be detailed later. Indeed, the marginal variance of the elements in $A$ is then not scaled only by $V$, but also $Q$. That scaling however is constant along the rows, and so is the scaling by the hyperparameter $s_f$ \eqref{eq:covfs}. We therefore suggest to simply use $s_f$ as tuning for the overall influence of the priors; letting $s_f \to \infty$ gives a flat prior, or, a non-regularized basis function expansion.

\subsubsection{Prior for noise covariances}
Apart from $f(\cdot)$, the  $\nx \times \nx$ noise covariance matrix $Q$ might also be unknown. We formulate the prior over $Q$ as an inverse Wishart ($\IW$, \citealt{Dawid:1981}) distribution. The $\IW$ distribution is a distribution over real-valued positive definite matrices, which puts prior mass on all positive definite matrices and is parametrized by its number of degrees of freedom $\ell > \nx - 1$ and an $\nx \times \nx$ positive definite scale matrix $\Lambda$. The density is defined as
\begin{equation}
\mathcal{IW}({Q}\mid\ell,{\Lambda}) = \frac{|{\Lambda}|^{\ell/2}|{Q}|^{-(\nx+\ell+1)/2}}{2^{\ell \nx/2}\Gamma_{\nx}(\ell/2)}\exp\left(-\frac{1}{2}\tra{{Q}^{-1}{\Lambda}}\right),\label{eq:IWpdf}
\end{equation}
where $\Gamma_{\nx}(\cdot)$ is the multivariate gamma function. The mode of the $\IW$ distribution is $\frac{\Lambda}{\ell+\nx+1}$. It is a common choice as a prior for covariance matrices due to its properties (e.g., \citealt{WSL+:2012,SWG:2014}). When the $\MN$ distribution~\eqref{eq:MNpdf} is combined with the $\IW$ distribution~\eqref{eq:IWpdf} we obtain the $\MNIW$ distribution, with the following hierarchical structure
\begin{equation}
\MNIW(A,Q\mid M, V, \Lambda, \ell) = \MN(A\mid M, Q, V) \IW(Q\mid \ell, \Lambda).\label{eq:MNIW}
\end{equation}

The $\MNIW$ distribution provides a joint prior for the $A$ and $Q$ matrices, compactly parameterizing the prior scheme we have discussed, and is also the conjugate prior for our model, which will facilitate learning.

\begin{wrapfigure}[15]{r}{.45\linewidth}
	\begin{center}
		\vspace{-2em}
		\includegraphics[width=\linewidth]{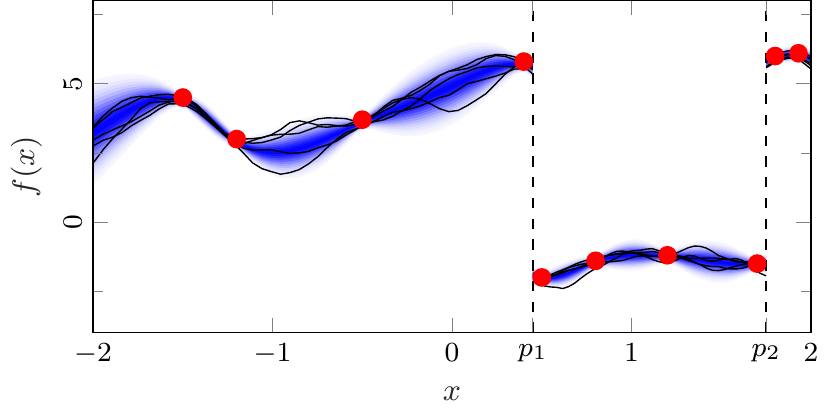}
		
		\caption{The idea of a piecewise GP: the interval $[-2,-2]$ is divided by $n_p = 2$ discontinuity points $p_1$ and $p_2$, and a GP is used to model a function on each of these segments, independently of the other segments. For practical use, the learning algorithm have to be able to also infer the discontinuity points from data.
		}
		\label{fig:pwGP}
	\end{center}
\end{wrapfigure}

\subsubsection{Discontinuous functions: Sparse singularities}\label{sec:model:sing}

The proposed choice of basis functions and priors is encoding a smoothness assumption of $f(\cdot)$. However, as discussed by \citet{JHB+:1995} and motivated by Example~\ref{sec:experiment:water}, there are situations where it is relevant to assume that $f(\cdot)$ is smooth \textit{except at a few points}. Instead of assuming an (approximate) GP prior for $f(\cdot)$ on the entire interval $[-L,L]$ we therefore suggest to divide $[-L,L]$ into a number $n_p$ of segments, and then assume an individual GP prior for each segment $[p_{i},p_{i+1}]$, independent of all other segments, as illustrated in Figure~\ref{fig:pwGP}. The number of segments and the discontinuity points dividing them need to be learned from data, and an important prior is how the discontinuity points are distributed, i.e., the number $n_p$ (e.g., geometrically distributed) and their locations $\{p_i\}_{i=1}^{n_p}$ (e.g., uniformly distributed).

\subsection{Model summary}
We will now summarize the proposed model. To avoid notational clutter, we omit $u_t$ as well as the observation function~\eqref{eq:ssm:b}:
\begin{subequations}\label{eq:fmodel}
\begin{align}
	x_{t+1} &= \sum_{i=0}^{n_p} A_i \phivec(x_t) \Inda{p_i \leq x_t < p_{i+1}} + v_t,\label{eq:fmodel:tr} \\
	v_t &\sim \N(0,Q),
\end{align}
with priors
\begin{align}
	[A_i,Q_i] \sim&~ \MNIW(0,V,\ell,\Lambda),~i=0, \dots, n_p,\\
	n_p, \{p_i\}_{i=1}^{n_p} \sim&~ \text{arbitrary prior},
\end{align}
\end{subequations}
where $\mathbbm{1}$ is the indicator function parameterizing the piecewise GP, and $\phivec(x_t)$ was defined in~\eqref{eq:ss2}.
If the dynamical behavior of the data is close-to-linear, and a fairly accurate linear model is already available, this can be incorporated by adding the known linear function to the right hand side of \eqref{eq:fmodel:tr}.

A good user practice is to sample parameters from the priors and simulate the model with those parameters, as a sanity check before entering the learning phase. Such a habit can also be fruitful for understanding what the prior assumptions mean in terms of dynamical behavior. There are standard routines for sampling from the $\MN$ as well as the $\IW$ distribution.

The suggested model can also be tailored if more prior knowledge is present, such as a physical relationship between two certain state variables. The suggested model can then be used to learn only the unknown part, as briefly illustrated by \citet[Example IV.B]{SSS+:2015}.

\section{Learning}\label{sec:learning}
We now have a state-space model with a (potentially large) number of unknown parameters
\begin{equation}
\theta \triangleq \Big\{\{A_i,Q_i\}_{i=0}^{n_p},n_p,\{p_i\}_{i=1}^{n_p}\Big\},
\end{equation}
all with priors. ($g(\cdot)$ is still assumed to be known, but the extension follows analogously.) Learning the parameters is a quite general problem, and several learning strategies proposed in the literature are (partially) applicable, including optimization \citep{PLS+:2010}, EM with extended Kalman filtering \citep{GR:1999} or sigma point filters \citep{KSS:2015}, and particle Metropolis-Hastings \citep{TDM:2015}. We use another sequential Monte Carlo-based learning strategy, namely particle Gibbs with ancestor sampling (PGAS, \citealt{LJS:2014}). PGAS allows us to take advantage of the fact that our proposed model~\eqref{eq:ss2} is linear in $A$ (given $x_t$), at the same time as it has desirable theoretical properties.

\subsection{Sequential Monte Carlo for system identification}

Sequential Monte Carlo (SMC) methods have emerged as a tool for learning parameters in state-space models \citep{SLD+:2015,KDS+:2015}. At the very core when using SMC for system identification is the particle filter \citep{DJ:2011}, which provides a numerical solution to the state filtering problem, i.e., finding $p(x_t\mid y_{1:t})$. The particle filter propagates a set of weighted samples, particles, $\{x_t^i, \omega^i_t\}_{i=1}^N$ in the state-space model, approximating the filtering density by the empirical distribution $\widehat{p}(x_t\mid y_{1:t}) = \sum_{i=1}^N \omega^i_t \delta_{x_t^i}(x_t)$ for each $t$. Algorithmically, it amounts to iteratively weighting the particles with respect to the measurement $y_t$, resample among them, and thereafter propagate the resampled particles to the next time step $t+1$. The convergence properties of this scheme have been studied extensively (see references in \citet{DJ:2011}).

When using SMC methods for learning parameters, a key idea is to repeatedly infer the unknown states $x_{1:T}$ with a particle filter, and interleave this iteration with inference of the unknown parameters $\theta$, as follows:
\begin{equation}
\begin{aligned}
\text{I. }&\text{Use SMC to infer the states $x_{1:T}$ for given parameters $\theta$.} \\
\text{II. }&\text{Update the parameters $\theta$ to fit the states $x_{1:T}$ from the previous step.}
\end{aligned}\label{it}
\end{equation}
There are several details left to specify in this iteration, and we will pursue two approaches for updating $\theta$: one sample-based for exploring the full posterior $p(\theta|y_{1:T})$, and one EM-based for finding the maximum mode of the posterior, or equivalently, a regularized maximum likelihood estimate. Both alternatives will utilize the linear-in-parameter structure of the model~\eqref{eq:fmodel}, and use the Markov kernel PGAS \citep{LJS:2014} to handle the states in Step I of~\eqref{it}.

The PGAS Markov kernel resembles a standard particle filter, but has one of its state-space trajectories fixed. It is outlined by Algorithm~\ref{alg:CPFAS}, and is a procedure to asymptotically produce samples from $p(x_{1:T}\mid y_{1:T},\theta)$, if repeated iteratively in a Markov chain Monte Carlo (MCMC, \citealt{RC:2004}) fashion.

\begin{algorithm}[t]
	\caption{PGAS Markov kernel.}
	\label{alg:CPFAS}
	\begin{algorithmic}[1]\small
		\Require Trajectory ${x}_{1:T}[k]$, number of particles $N$, known state-space model ($f$, $g$, $Q$, $R$).
		\Ensure Trajectory ${x}_{1:T}[k+1]$
		\State Sample ${x}_1^\iip \sim p({x}_1)$, for $i = 1,\dots,N-1$.
		\State Set ${x}_1^N = {x}_1[k]$.
		\For{$t = 1$ to $T$}
		\State ~Set $\pfw_t^\iip = \N\left(y_t\mid {g}({x}_t^\iip),R\right)$, for $i = 1, \dots, N$.
		\State ~\label{alg:CPFAS:resampling}Sample $\anc_t^\iip$ with $\Prb{\anc_t^\iip=j} \propto \pfw_{t}^\jip$, for $i = 1, \dots, N-1$.
		\State ~Sample ${x}_{t+1}^\iip \sim \N\left({f}({x}_t^{\anc_t^\iip}),Q\right)$, for $i = 1, \dots, N-1$.
		\State ~Set ${x}_{t+1}^N = {x}_{t+1}[k]$.
		\State ~Sample $\anc_t^\Nip$ w. $\Prb{\anc_t^\Nip = j} \propto \pfw_{t}^\jip\N \left({{x}_{t+1}^\Nip}\mid{{f}({x}_t^\jip)},Q\right)$.
		\State ~Set ${x}_{1:t+1}^\iip = \{{x}_{1:t}^{\anc_t^\iip},{x}_{t+1}^\iip\}$, for $i = 1, \dots, N$.
		\EndFor
		\State Sample $J$ with $\Prb{J=i} \propto \pfw_T^\iip$ and set ${x}_{1:T}[k+1] = {x}_{1:T}^J$.
	\end{algorithmic}
\end{algorithm}

\subsection{Parameter posterior}

The learning problem will be split into the iterative procedure~\eqref{it}. In this section, the focus is on a key to Step II of~\eqref{it}, namely the conditional distribution of $\theta$ given states $x_{1:T}$ and measurements $y_{1:T}$. By utilizing the Markovian structure of the state-space model, the density $p(x_{1:T},y_{1:T}\mid \theta)$ can be written as the product
\begin{equation}
p(x_{1:T},y_{1:T}\mid \theta) = p(x_1)\prod_{t=1}^{T-1}p(x_{t+1}\mid x_t, \theta)p(y_t\mid x_t)
= \underbrace{p(x_1)\prod_{t=1}^{T-1}p(x_{t+1}\mid x_t, \theta)}_{p(x_{1:T}\mid\theta)}\underbrace{\prod_{t=1}^{T}p(y_t\mid x_t)}_{p(y_{1:T}\mid x_{1:T})}\label{eq:xlik}.
\end{equation}
Since we assume that the observation function~\eqref{eq:ssm:b} is known, $p(y_t\mid x_t)$ is independent of $\theta$, which in turn means that~\eqref{eq:xlik} is proportional to $p(x_{1:T}\mid\theta)$. Further, we assume for now that $p(x_1)$ is also known, and therefore omit it. Let us consider the case without discontinuity points, $n_p = 0$. Since $v_t$ is assumed to be Gaussian, $p(x_{t+1}\mid x_t, u_t, \theta) =  \N(x_{t+1}\mid A\phivec(x_t,u_t),Q)$, we can with some algebraic manipulations \citep{GN:2005} write
\begin{equation}
\log p(x_{1:T}\mid A,Q) = -\tfrac{T\nx}{2}\log(2\pi) -\tfrac{T}{2} \log \det (Q) - \tfrac{1}{2}\tr \left\{ Q^{-1}\left(\Phi - A\Psi^\Transp - \Psi A^\Transp + A\Sigma A^\Transp\right)\right\},\label{eq:ll}
\end{equation}
with the (sufficient) statistics
\begin{subequations}\label{eq:stats}
\begin{equation}
\Phi = \sum_{t=1}^{T} x_{t+1}x_{t+1}^\Transp,\quad
\Psi = \sum_{t=1}^{T} x_{t+1}\phivec(x_t,u_t)^\Transp, \quad
\Sigma = \sum_{t=1}^{T} \phivec(x_t,u_t)\phivec(x_t,u_t)^\Transp.
\end{equation}
\end{subequations}
The density~\eqref{eq:ll} gives via Bayes' rule and the $\MNIW$ prior distribution for $A,Q$ from Section~\ref{sec:model}
\begin{equation}
\log p(A,Q) = \log p(A\mid Q) + \log p(Q) \propto 
-\tfrac{1}{2}(\nx+\ell+m+1)\log\det(Q)-\tfrac{1}{2}\mathrm{tr}\left\{Q^{-1}\left(\Lambda + AV^{-1}A^\Transp\right)\right\},\label{eq:prior}
\end{equation}
the posterior
\begin{multline}
\log p(A,Q\mid x_{1:t}) \propto \log p(x_{1:t}\mid A,Q) + \log p(A,Q) \propto 
 -\tfrac{1}{2}(\nx+T+\ell+m+1) \log \det Q \\ - \tfrac{1}{2}\tr \big\{Q^{-1}\big(\Lambda + \Phi - \Psi(\Sigma+V^{-1})^{-1}\Psi^\Transp +(A-\Psi(\Sigma+V^{-1})^{-1})Q^{-1}(A-\Psi(\Sigma+V^{-1})^{-1})^\Transp\big)\big\}.\label{eq:posterior}
\end{multline}
This expression will be key for learning: For the fully Bayesian case, we will recognize~\eqref{eq:posterior} as another $\MNIW$ distribution and sample from it, whereas we will maximize it when seeking a point estimate.

\textit{Remarks:} The expressions needed for an unknown observation function $g(\cdot)$ are completely analogous. The case with discontinuity points becomes essentially the same, but with individual $A_i,Q_i$ and statistics for each segment. If the right hand side of~\eqref{eq:fmodel:tr} also contains a known function $h(x_t)$, e.g., if the proposed model is used only to describe deviations from a known linear model, this can easily be taken care of by noting that now $p(x_{t+1}\mid x_t, u_t, \theta) =  \N(x_{t+1}-h(x_t)\mid A\phivec(x_t,u_t),Q)$, and thus compute the statistics~\eqref{eq:stats} for $(x_{t+1}-h(x_t))$ instead of $x_{t+1}$.

\subsection{Inferring the posterior---Bayesian learning}\label{sec:learning:bayesian}

There is no closed form expression for $p(\theta\mid y_{1:T})$, the distribution to infer in the Bayesian learning. We thus resort to a numerical approximation by drawing samples from $p(\theta,x_{1:T}\mid y_{1:T})$ using MCMC. (Alternative, variational methods could be used, akin to \citet{FCR:2014}). MCMC amounts to constructing a procedure for `walking around' in $\theta$-space in such a way that the steps $\dots, \theta[k], \theta[k+1], \dots$ eventually, for $k$ large enough, become samples from the distribution of interest.

Let us start in the case without discontinuity points, i.e., $n_p \equiv 0$. Since \eqref{eq:prior} is $\MNIW$, and \eqref{eq:ll} is a product of (multivariate) Gaussian distributions, \eqref{eq:posterior} is also an $\MNIW$ distribution \citep{WSL+:2012,Dawid:1981}. By identifying components in~\eqref{eq:posterior}, we conclude that
\begin{equation}
	p(\theta\mid x_{1:T}, y_{1:T}) = 
	\MNIW\big(A,Q\mid\Psi(\Sigma+V^{-1})^{-1},(\Sigma+V^{-1})^{-1},\Lambda + \Phi - \Psi(\Sigma+V^{-1})^{-1}\Psi^\Transp,\ell+Tn_x\big)\label{eq:mnpost}
\end{equation}
We now have~\eqref{eq:mnpost} for sampling $\theta$ given the states $x_{1:T}$ (cf. \eqref{it}, step II), and Algorithm~\ref{alg:CPFAS} for sampling the states $x_{1:T}$ given the model $\theta$ (cf. \eqref{it}, step I). This makes a particle Gibbs sampler \citep{ADH:2010}, cf.~\eqref{it}.


If there are discontinuity points to learn, i.e., $n_p$ is to be learned, we can do that by acknowledging the hierarchical structure of the model. For brevity, we denote $\{n_p, \{p_i\}_{i=1}^{n_p}\}$ by $\xi$, and $\{A_i,Q_i\}_{i=1}^{n_p}$ simply by $A,Q$. We suggest to first sample $\xi$ from $p(\xi\mid x_{1:T})$, and next sample $A,Q$ from $p(A,Q \mid x_{1:T},\xi)$. The distribution for sampling $A,Q$ is the $\MNIW$ distribution~\eqref{eq:mnpost}, but conditional on data only in the relevant segment. The other distribution, $p(\xi\mid x_{1:T})$, is trickier to sample from. We suggest to use a Metropolis-within-Gibbs step \citep{Muller:1991}, which means that we first sample $\xi^*$ from a proposal $q(\xi^*\mid\xi[k])$ (e.g., a random walk), and then accept it as $\xi[k\!+\!1]$ with probability $\min\left(1,\frac{p(\xi^*\mid x_{1:T})}{p(\xi[k]\mid x_{1:T})}\frac{q(\xi[k]\mid\xi[k])}{q(\xi^*\mid\xi[k])}\right)$, and otherwise just set $\xi[k\!+\!1] = \xi[k]$. Thus we need to evaluate $p(\xi^*\mid x_{1:T}) \propto p(x_{1:T}\mid\xi^*)p(\xi^*)$. The prior $p(\xi^*)$ is chosen by the user. The density $p(x_{1:T}\mid \xi)$ can be evaluated using the expression (see Appendix~\ref{app:marglik})
\begin{equation}
	p(x_{1:T}\mid \xi) = \prod_{i=0}^{n_p}\frac{2^{n_x T_i/2}}{(2\pi)^{T_i/2}}
	\frac{\Gamma_{\nx}(\tfrac{l+N}{2})}{\Gamma_{\nx}(\tfrac{l}{2})} \frac{|V^{-1}|^{n_x/2}}{|\Sigma_i+V^{-1}|^{n_x/2}} 
	\times
	\frac{|\Lambda|^{l/2}}{|\Lambda+\Phi_i+\Psi_i(\Sigma_i+V^{-1})^{-1}\Psi_i^\Transp|^{\frac{l+N}{2}}}\label{eq:marglik}
\end{equation}
where $\Phi_i$ etc. denotes the statistics~\eqref{eq:stats} restricted to the corresponding segment, and $T_i$ is the number of data points in segment $i$ ($\sum_i T_i = T$). The suggested Bayesian learning procedure is summarized in Algorithm~\ref{alg:gibbs}.

\begin{algorithm}[t]
	\caption{Bayesian learning of \eqref{eq:fmodel}}
	\begin{algorithmic}[1]\small
		\Require Data $y_{1:T}$, priors on $A,Q$ and ${\xi}$.
		\Ensure $K$ MCMC-samples with $p({x}_{1:T},A,Q,\xi\mid{y}_{1:T})$ as invariant distribution.
		\State Initialize ${A}[0],{Q}[0], \xi[0]$.
		\For{$k = 0$ to $K$}
		\State\label{alg:gibbs:x}
		Sample ${{x}_{1:T}[k\!+\!1]}
		\boldsymbol{\:\big\vert\:}{A}[k],{Q}[k],\xi[k]$\hfill Algorithm~\ref{alg:CPFAS}
		\State\label{alg:gibbs:q}
		Sample \phantom{${x}_{1:T}[k\!+\!1]$}$\mathllap{{\xi}[k\!+\!1]}
		\boldsymbol{\:\big\vert\:}{x}_{1:T}[k\!+\!1]$\hfill Section~\ref{sec:learning:bayesian}
		\State\label{alg:gibbs:q}
		Sample \phantom{${x}_{1:T}[k\!+\!1]$}$\mathllap{{Q}[k\!+\!1]}
		\boldsymbol{\:\big\vert\:}\xi[k\!+\!1],{x}_{1:T}[k\!+\!1]$\hfill by  \eqref{eq:mnpost}
		\State\label{alg:gibbs:a}
		Sample \phantom{${x}_{1:T}[k\!+\!1]$}$\mathllap{{A}[k\!+\!1]}
		\boldsymbol{\:\big\vert\:}{Q}[k\!+\!1], \xi[k\!+\!1], {x}_{1:T}[k\!+\!1]$\hfill by \eqref{eq:mnpost}
		\EndFor
	\end{algorithmic}
	\label{alg:gibbs}
\end{algorithm}

Our proposed algorithm can be seen as a combination of a collapsed Gibbs sampler and Metropolis-within-Gibbs, a combination which requires some attention to be correct \citep{DJ:2014}, see Appendix~\ref{app:mh-w-g} for details in our case. If the hyperparameters parameterizing $V$ and/or the initial states are unknown, it can be included by extending Algorithm~\ref{alg:gibbs} with extra Metropolis-within-Gibbs steps (see \citet{SSS+:2016} for details).

\newpage

\subsection{Regularized maximum likelihood}

A widely used alternative to Bayesian learning is to find a \textit{point estimate} of $\theta$ maximizing the likelihood of the training data $p(y_{1:T}\mid \theta)$, i.e., \textit{maximum likelihood}. However, if a very flexible model is used, some kind of mechanism is needed to prevent the model from overfit to training data. We will therefore use the priors from Section~\ref{sec:model} as regularization for the maximum likelihood estimation, which can also be understood as seeking the maximum mode of the posterior. We will only treat the case with no discontinuity points, as the case with discontinuity points does not allow for closed form maximization, but requires numerical optimization tools, and we therefore suggest Bayesian learning for that case instead.

The learning will build on the particle stochastic approximation EM (PSAEM) method proposed by \citet{Lindsten:2013}, which uses a stochastic approximation of the EM scheme \citep{DLR:1977,DLM:1999,KL:2004}. EM addresses maximum likelihood estimation in problems with latent variables. For system identification, EM can be applied by taking the states $x_{1:T}$ as the latent variables, (\citet{GR:1999}; another alternative would be to take the noise sequence $v_{1:T}$ as the latent variables, \citet{UWM+:2015}).
\begin{subequations}\label{eq:em}
The EM algorithm then amounts to iteratively (cf.~\eqref{it}) computing the expectation (E-step)
\begin{equation}
\emQ(\theta,\theta[k]) = \Expb{x_{1:T}}{\log p(\theta\mid x_{1:T},y_{1:T})\mid y_{1:T}, \theta[k]},\label{eq:em:q}
\end{equation}
and updating $\theta$ in the maximization (M-step) by solving
\begin{equation}
\theta[k\!+\!1] = \arg\max_\theta \emQ(\theta,\theta[k]),\label{eq:em:max}
\end{equation}
\end{subequations}
In the standard formulation, $\emQ$ is usually computed with respect to the joint likelihood density for $x_{1:T}$ and $y_{1:T}$. To incorporate the prior (our regularization), we may consider the prior as an additional observation of $\theta$, and we have thus replaced~\eqref{eq:ll} by~\eqref{eq:posterior} in~$\emQ$. Following \citet{GN:2005}, the solution in the M-step is found as follows: Since $Q^{-1}$ is positive definite, the quadratic form in~\eqref{eq:posterior} is maximized by
\begin{subequations}\label{eq:em:maxs}
\begin{equation}
A = \Phi(\Sigma + V^{-1}).
\end{equation}
Next, substituting this into~\eqref{eq:posterior}, the maximizing $Q$ is
\begin{equation}
Q = \tfrac{1}{\nx+T\nx+\ell+m+1}\left(\Lambda + \Phi - \Psi(\Sigma + V^{-1})^{-1}\Psi\right).
\end{equation}
\end{subequations}
We thus have solved the M-step exactly. To compute the expectation in the E-step, approximations are needed. For this, a particle smoother \citep{LS:2013} could be used, which would give a learning strategy in the flavor of \citet{SWN:2011}. The computational load of a particle smoother is, however, unfavorable, and PSAEM uses Algorithm~\ref{alg:CPFAS} instead.

PSAEM also replaces and replace the $\emQ$-function~\eqref{eq:em:q} with a Robbins-Monro stochastic approximation of $\emQ$,
\begin{equation}
\saemQ_k(\theta) = (1-\gamma_k)\saemQ_{k-1}(\theta) + \gamma_k \log p(\theta\mid x_{1:T}[k],y_{1:T}),\label{eq:saem:q}
\end{equation}
where $\{\gamma_k\}_{k\geq 1}$ is a decreasing sequence of positive step sizes, with $\gamma_1 = 1$, $\sum_k \gamma_k = \infty$ and $\sum_k \gamma_k^2 < \infty$. I.e., $\gamma_k$ should be chosen such that $k^{-1} \leq \gamma_k < k^{-0.5}$ holds up to proportionality, and the choice $\gamma_k = k^{-2/3}$ has been suggested in the literature \citep[Section 5.1]{DLM:1999}. Here, $x_{1:T}[k]$ is a sample from an ergodic Markov kernel with $p(x_{1:T}\mid y_{1:T},\theta)$ as its invariant distribution, i.e., Algorithm~\ref{alg:CPFAS}. At a first glance, the complexity of $\saemQ_k(\theta)$ appears to grow with $k$ because of its iterative definition. However, since $p(x_{1:T},y_{1:T}\mid\theta)$ belongs to the exponential family, we can write
\begin{equation}
p(x_{1:T}[k],y_{1:T}\mid\theta) = h(x_{1:T}[k],y_{1:T})c(\theta)\exp\left(\eta^\Transp(\theta)t[k]\right),
\end{equation}
where $t[k]$ is the statistics~\eqref{eq:stats} of $\{x_{1:T}[k],y_{1:T}\}$. The stochastic approximation $\saemQ_k(\theta)$~\eqref{eq:saem:q} thus becomes
\begin{equation}
\saemQ_k(\theta) \propto \log p(\theta) + \log c(\theta) + \eta^\Transp(\theta)\left(\gamma_k t[k] + (1-\gamma_k)\gamma_{k\text{-}1}t[k-1] + \dots\right).
\end{equation}
Now, we note that if keeping track of the statistics $\gamma_k t[k] + \gamma_{k\text{-}1}t[k\text{-}1] + \dots$, the complexity of $\saemQ$ does not grow with $k$. We therefore introduce the following iterative update of the statistics
\begin{subequations}\label{eq:saem:stats}
\begin{align}
\Phi_k   &= (1-\gamma_k)\Phi_{k-1}   + \gamma_k\Phi(x_{1:T}[k]),\\
\Psi_k   &= (1-\gamma_k)\Psi_{k-1}   + \gamma_k\Psi(x_{1:T}[k]),\\
\Sigma_k &= (1-\gamma_k)\Sigma_{k-1} + \gamma_k\Sigma(x_{1:T}[k]),
\end{align}
\end{subequations}
where $\Phi(x_{1:T}[k])$ refers to \eqref{eq:stats}, etc. With this parametrization, we obtain $\arg\max_\theta \saemQ_k(\theta)$ as the solutions for the vanilla EM case by just replacing $\Phi$ by $\Phi_k$, etc., in~\eqref{eq:em:maxs}. Algorithm~\ref{alg:PSAEM_id} summarizes.

\begin{algorithm}[t]
	\caption{Regularized maximum likelihood}
	\begin{algorithmic}[1]\small
		\State Initialize $\theta{[1]}$.
		\For{$k > 0$}
		\State Sample $x_{1:T}[k]$ by Algorithm~\ref{alg:CPFAS} with parameters $\theta[k]$.
		\State Compute and update the statistics of $x_{1:T}[k]$~(\ref{eq:stats}, \ref{eq:saem:stats}).
		\State Compute $\theta{[k\!+\!1]} = \arg\max_\theta \saemQ(\theta)$~\eqref{eq:em:maxs}.
		\EndFor 
	\end{algorithmic}
	\label{alg:PSAEM_id}
\end{algorithm}

\subsection{Convergence and consistency}

We have proposed two algorithms for learning the model introduced in Section~\ref{sec:model}. The Bayesian learning, Algorithm~\ref{alg:gibbs}, will by construction (as detailed in Appendix~\ref{app:mh-w-g}) asymptotically provide samples from the true posterior density $p(\theta\mid y_{1:T})$ \citep{ADH:2010}. However, no guarantees regarding the length of the burn-in period can be given, which is the case for all MCMC methods, but the numerical comparisons in \citet{SSS+:2016} and in Section~\ref{sec:experiment:toy} suggest that the proposed Gibbs scheme is efficient compared to its state-of-the-art alternatives. The regularized maximum likelihood learning, Algorithm~\ref{alg:PSAEM_id}, can be shown to converge under additional assumptions \citep{Lindsten:2013,KL:2004} to a stationary point of $p(\theta|y_{1:T})$, however not necessarily a global maximum. The literature on PSAEM is not (yet) very rich, and the technical details regarding the additional assumptions remains to be settled, but we have not experienced any problems of non-convergence in practice.

\subsection{Initialization}
The convergence of Algorithm~\ref{alg:gibbs} is not relying on the initialization, but the burn-in period can nevertheless be reduced. One useful idea by \citet{PLS+:2010} is thus to start with a linear model, which can be obtained using classical methods. To avoid Algorithm~\ref{alg:PSAEM_id} from converging to a poor local minimum, Algorithm~\ref{alg:gibbs} can first be run to explore the `landscape' and from that, a promising point for initialization of Algorithm~\ref{alg:PSAEM_id} can be chosen.

For convenience, we assumed the distribution of the initial states, $p(x_1)$, to be known. This is perhaps not realistic, but its influence is minor in many cases. If needed, they can be included in Algorithm~\ref{alg:gibbs} by an additional Metropolis-within-Gibbs step, and in Algorithm~\ref{alg:PSAEM_id} by including them in~\eqref{eq:posterior} and use numerical optimization tools.

\clearpage

\begin{wrapfigure}{r}{.46\linewidth}
	\hspace*{\fill}
	\begin{subfigure}[b]{\linewidth}
		\centering
		\includegraphics[width=.9\linewidth]{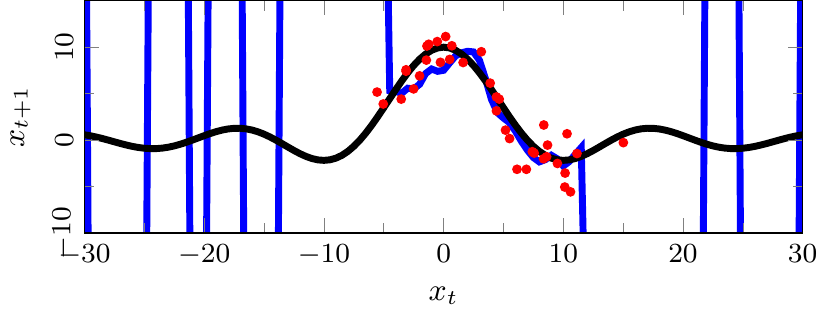}
		\vspace{-.7em}
		\caption{Maximum likelihood estimation of our proposed model, \textit{without} regularization; a useless model.}
		\label{fig:simple:1}
	\end{subfigure}
\vspace{1.5em}
	\begin{subfigure}[b]{\linewidth}
		\centering
		\includegraphics[width=.9\linewidth]{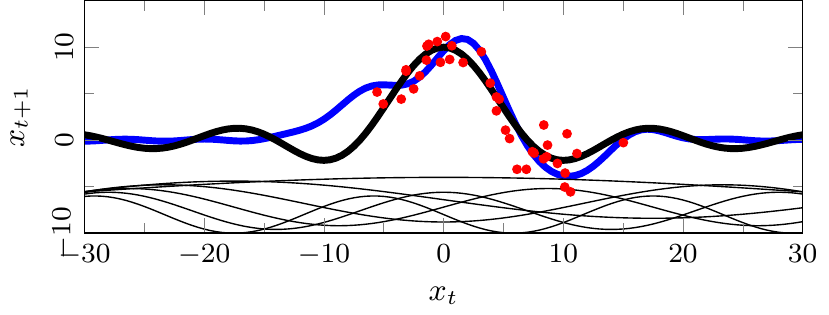}
		\vspace{-.7em}
		\caption{Maximum likelihood estimation of our proposed model, \textit{with} regularization. A subset of the $m=40$ basis functions used are sketched at the bottom. Computation time: 12 s.}
		\label{fig:simple:2}
	\end{subfigure}
\vspace{.5em}
	\begin{subfigure}[b]{\linewidth}
		\centering
		\includegraphics[width=.9\linewidth]{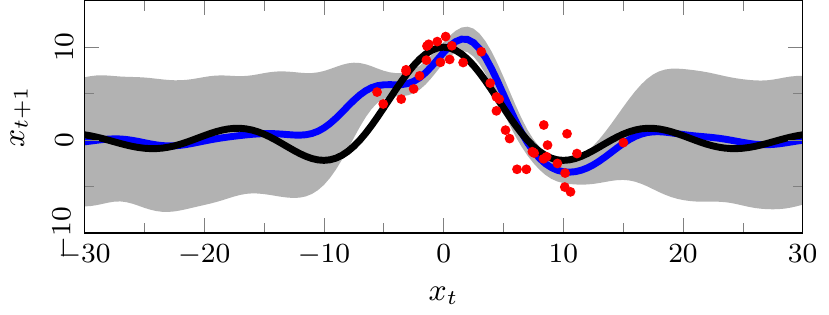}
		\vspace{-.7em}
		\caption{Bayesian learning of our proposed model, i.e., the entire posterior is explored. Computation time: 12 s.}
		\label{fig:simple:3}
	\end{subfigure}
\vspace{.5em}
	\begin{subfigure}[b]{\linewidth}
		\centering
		\includegraphics[width=.9\linewidth]{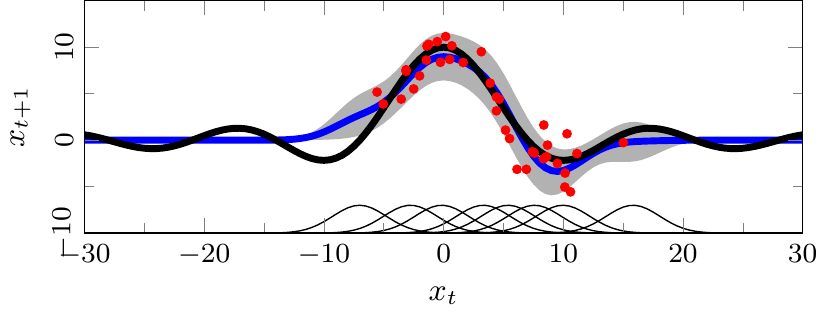}
		\vspace{-.7em}
		\caption{Posterior distribution for the basis functions (sketched at the bottom) used by \citet{TDM:2015}, but Algorithm~\ref{alg:gibbs} for learning. Computation time: 9 s.}
		\label{fig:simple:4}
	\end{subfigure}
\vspace{.5em}
	\begin{subfigure}[b]{\linewidth}
		\centering
		\includegraphics[width=.9\linewidth]{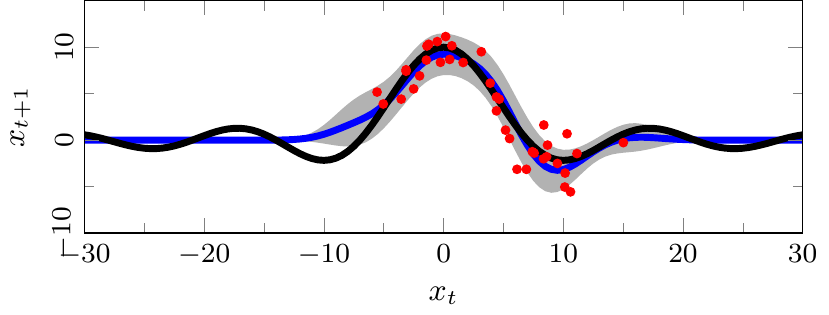}
		\vspace{-.7em}
		\caption{The method presented by \citet{TDM:2015}, using Metropolis-Hastings for learning. Computation time: 32 s.}
		\label{fig:simple:5}
	\end{subfigure}
\vspace{.5em}
	\begin{subfigure}[b]{\linewidth}
		\includegraphics[width=\linewidth]{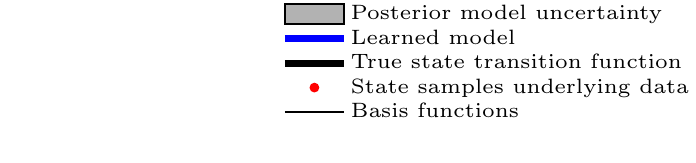}
		\vspace{-3.5em}
	\end{subfigure}
	\caption{True function (black), states underlying the data (red) and learned model (blue, gray) for the example in Section~\ref{sec:experiment:toy}.}
	\label{fig:simple}
\end{wrapfigure}

\section{Experiments}\label{sec:experiment}

We will give three numerical examples: a toy example, a classic benchmark, and thereafter a real data set from two cascaded water tanks. Matlab code for all examples is available via the first authors homepage.

\subsection{A first toy example}\label{sec:experiment:toy}

Consider the following example from \citet{TDM:2015},
\begin{subequations}
\begin{align}
x_{t+1} &= 10 \text{sinc}\left(\frac{x_t}{7}\right) + v_t, & v_t\sim\N(0,4),\\
y_t &= x_t + e_t, & e_t\sim\N(0,4).
\end{align}
\end{subequations}
We generate $T = 40$ observations, and the challenge is to learn $f(\cdot)$, when $g(\cdot)$ and the noise variances are known. Note that even though $g(\cdot)$ is known, $y$ is still corrupted by a non-negligible amount of noise.

In Figure~\ref{fig:simple} (a) we illustrate the performance of our proposed model using $m=40$ basis functions on the form~\eqref{eq:bf} when Algorithm~\ref{alg:PSAEM_id} is used \textit{without} regularization. This gives a nonsense result that is overfitted to data, since $m=40$ offers too much flexibility for this example. When a GP-inspired prior from an exponentiated quadratic covariance function~\eqref{eq:eq} with length scale $\ell = 3$ and $s_f = 50$ is considered, we obtain (b), that is far more useful and follows the true function rather well in regions were data is present. We conclude that we do \textit{not} need to choose $m$ carefully, but can rely on the priors for regularization. In (c), we use the same prior and explore the full posterior by Algorithm~\ref{alg:gibbs}, obtaining information about uncertainty as a part of the learned model (illustrated by the a posteriori credibility interval), in particular in regions where no data is present.

In the next figure, (d), we replace the set of $m=40$ basis functions on the form~\eqref{eq:bf} with $8$ Gaussian kernels to reconstruct the model proposed by \citet{TDM:2015}. As clarified by \citet{Tobar:2016}, the prior on the coefficients is a Gaussian distribution inspired by a GP, which makes a close connection to out work. We use Algorithm~\ref{alg:gibbs} for learning also in (d) (which is possible thanks to the Gaussian prior). In (e), on the contrary, the learning algorithm from \citet{TDM:2015}, Metropolis-Hastings, is used, requiring more computation time. \citet{TDM:2015} spend a considerable effort to pre-process the data and carefully distribute the Gaussian kernels in the state space, see the bottom of (d).

\clearpage
\subsection{Narendra-Li benchmark}

The example introduced by \cite{NL:1996} has become a benchmark for nonlinear system identification, e.g., \citealt{MathWorks:2015a,PLL:2009,RNL:2005,Stenman:1999,WWJ+:2007,XHW:2009}. The benchmark is defined by the model
\begin{subequations}
	\begin{align}
	x^1_{t+1} =& \left(\tfrac{x^1_t}{1+(x_t^1)^2}+1\right)\sin(x_t^2),\\
	x^2_{t+1} =& x^2_t\cos(x^2_t) + x^1_t\exp\left(-\tfrac{(x_t^1)^2+(x_t^2)^2}{8}\right) 
	+ \tfrac{(u_t)^3}{1+(u_t)^2+0.5\cos(x_t^1+x_t^2)}, \\
	y_{t} =& \tfrac{x_t^1}{1+0.5\sin(x_t^2)} + \tfrac{x_t^2}{1+0.5\sin(x_t^1)},\label{eq:nl:g}
	\end{align}
\end{subequations}
where $x_t = [x_t^1~x_t^2]^\Transp$. The training data (only input-output data) is obtained with an input sequence sampled uniformly and iid from the interval $[-2.5,2.5]$. The input data for the test data is $u_t = \sin(2\pi t/10) + \sin(2\pi t/25)$.

According to \citet[p. 369]{NL:1996}, it `does not correspond to any real physical system and is deliberately chosen to be complex and distinctly nonlinear'. The original formulation is somewhat extreme, with no noise and $T=500\,000$ data samples for learning. In the work by \citet{Stenman:1999}, a white Gaussian measurement noise with variance $0.1$ is added to the training data, and less data is used for learning. We apply Algorithm~\ref{alg:gibbs} with a second order state-space model, $n_p = 0$, and a known, linear $g(\cdot)$. (Even though the data is generated with a nonlinear $g(\cdot)$, it turn out this will give a satisfactory performance.) We use $7$ basis functions per dimension (i.e., $686$ coefficients $w^\ji$ to learn in total) on the form~\eqref{eq:mdbf}, with prior from the covariance function~\eqref{eq:eq} with length scale $\ell = 1$.

For the original case without any noise, but using only $T=500$ data points, a root mean square error (RMSE) for the simulation of $0.039$ is obtained.
Our result is in contrast to the significantly bigger simulation errors by \citet{NL:1996}, although they use $1\,000$ times as many data points. For the more interesting case \emph{with} measurement noise in the training data, we achieve a result almost the same as for the noise-free data. We compare to some previous results reported in the literature ($T$ is the number of data samples in the training data):

\begin{center}
{\small
\def\arraystretch{1}
\begin{tabular}{ l | l | l }
	Reference & RMSE & $T$ \\ \hline
	This paper & $\boldsymbol{0.06}$* & $2\,000$ \\
	\citet{RNL:2005} & $0.43$ & $50\,000$ \\
	\citet{Stenman:1999} & $0.46$ & $50\,000$ \\
	\citet{XHW:2009} (AHH) & $0.31$ & $2\,000$ \\
	\citet{XHW:2009} (MARS) & $0.49$ & $2\,000$ \\
\end{tabular}
}\\
*The number is averaged over 10 realizations
\end{center}

It is clear that the proposed model is capable enough to well describe the system behavior.

\subsection{Water tank data}\label{sec:experiment:water}

We consider the data sets provided by \citet{SMW+:2015}, collected from a physical system consisting of two cascaded water tanks, where the outlet of the first tank goes into the second one. A training and a test data set is provided, both with $1024$ data samples. The input $u$ (voltage) governs the inflow to the first tank, and the output $y$ (voltage) is the measured water level in the second tank. This is a well-studied system (e.g., \citealt{WS:2013}), but a peculiarity in this data set is the presence of overflow, both in the first and the second tank. When the first tank overflows, it goes only partly into the second tank.

We apply our proposed model, with a two dimensional state space. The following structure is used:
\begin{subequations}
	\begin{align}
	x^1_{t+1} &= f^1(x^1_t,u_t) + v_t^1,\\
	x^2_{t+1} &= f^2(x^1_t,x^2_t,u_t) + v_t^2,\\
	y_t &= x_t^2 + e_t.
	\end{align}
\end{subequations}
It is surprisingly hard to perform better than linear models in this problem, perhaps because of the close-to-linear dynamics in most regimes, in combination with the non-smooth overflow events. This calls for discontinuity points to be used. Since we can identify the overflow level in the second tank directly in the data, we fix a discontinuity point at $x^2 = 10$  for $f^2(\cdot)$, and learn the discontinuity points for $f^1(\cdot)$. Our physical intuition about the water tanks is a close-to-linear behavior in most regimes, apart from the overflow events, and we thus use the covariance function~\eqref{eq:eq} with a rather long length scale $\ell=3$ as prior. We also limit the number of basis functions to $5$ per dimension for computational reasons (in total, there are $150$ coefficients $w^\ji$ to learn).

\begin{wrapfigure}{r}{.5\linewidth}
	\begin{center}
		\includegraphics[width=\linewidth]{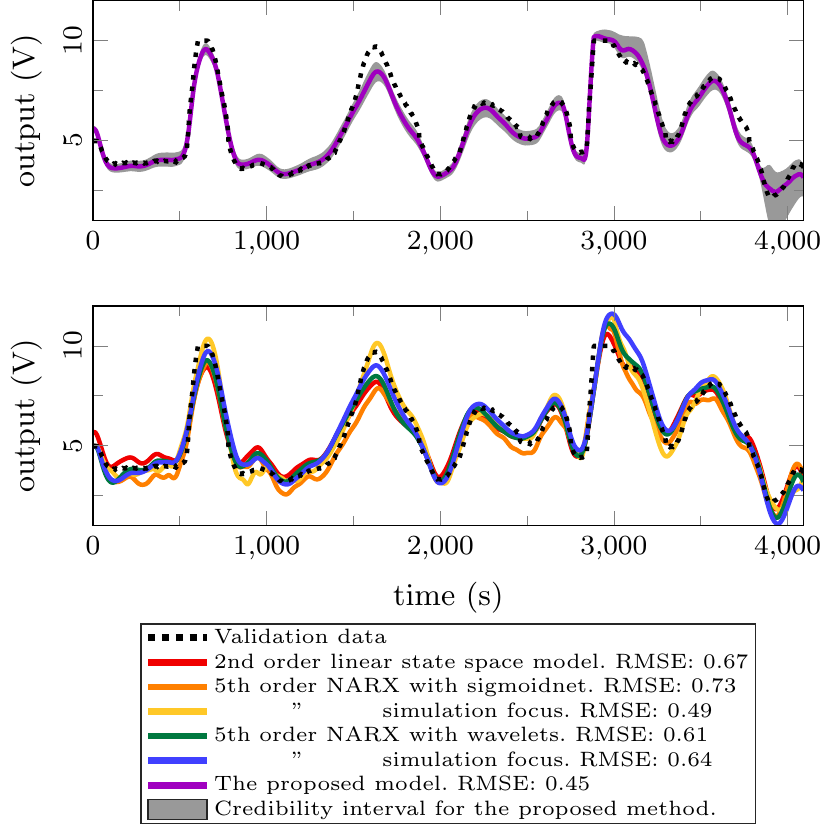}
		\caption{The simulated and true output for the test data in the water tank experiment (Section~\ref{sec:experiment:water}). The order of the NARX models refers to the number of regressors in $u$ and $y$.}
		\label{fig:wt_1}
	\end{center}
\end{wrapfigure}

Algorithm~\eqref{alg:gibbs} is used to sample from the model posterior. We use all samples to simulate the test output from the test input for each model to represent a posterior for the test data output, and compute the RMSE for the difference between the posterior mode and the true test output. A comparison to nonlinear ARX-models (NARX, \citealt{Ljung:1999}) is also made in Figure~\ref{fig:wt_1}. It is particularly interesting to note how the different models handle the overflow around time $3\,000$ in the test data. We have tried to select the most favorable NARX configurations, and when finding their parameters by maximizing their likelihood (which is equivalent to minimizing their 1-step-ahead prediction, \citealt{Ljung:1999}), the best NARX model is performing approximately $35\%$ worse (in terms of RMSE) than our proposed model. When instead learning the NARX models with `simulation focus', i.e., minimizing their simulation error on the training data, their RMSE decreases, and approaches almost the one of our model for one of the models\footnote{Since the corresponding change in learning objective is not available to our model, this comparison might only offer partial insight. It would, however, be an interesting direction for further research to implement learning with `simulation focus' in the Bayesian framework.}. While the different settings in the NARX models have a large impact on the performance, and therefore a trial-and-error approach is needed for the user to determine satisfactory settings, our approach offers a more systematic way to encode the physical knowledge at hand into the modeling process, and achieves a competitive performance.

\section{Conclusions and further work}

During the recent years, there has been a rapid development of powerful parameter estimation tools for state-space models. These methods allows for learning in complex and extremely flexible models, and this paper is a response to the situation when the learning algorithm is able to learn a state-space model more complex than the information contained in the training data (cf. Figure~\ref{fig:simple:1}). For this purpose, we have in the spirit of \citet{Peterka:1981} chosen to formulate GP-inspired priors for a basis function expansion, in order to `softly' tune its complexity and flexibility in a way that hopefully resonates with the users intuition. In this sense, our work resembles the recent work in the machine learning community on using GPs for learning dynamical models (see, e.g., \citealt{Frigola:2015,BSW+:2016,MDD+:2015}). However, not previously well explored in the context of dynamical systems, is the combination of discontinuities and the smooth GP. We have also tailored efficient learning algorithms for the model, both for inferring the full posterior, and finding a point estimate.

It is a rather hard task to make a sensible comparison between our \emph{model-focused approach}, and approaches which provide a general-purpose black-box learning algorithm with very few user choices. Because of their different nature, we do not see any ground to claim superiority of one approach over another. In the light of the promising experimental results, however, we believe this model-focused perspective can provide additional insight into the nonlinear system identification problem. There is certainly more to be done and understand when it comes to this approach, in particular concerning the formulation of priors.

We have proposed an algorithm for Bayesian learning of our model, which renders $K$ samples of the parameter posterior, representing a \textit{distribution} over models. A relevant question is then how to compactly represent and use these samples to efficiently make predictions. Many control design methods provide performance guarantees for a perfectly known model. An interesting topic would hence be to incorporate model \textit{uncertainty} (as provided by the posterior) into control design and provide probabilistic guarantees, such that performance requirements are fulfilled with, e.g., $95\%$ probability.

\normalsize
\appendix
\section{Appendix: Technical details}    
\subsection{Derivation of~\eqref{eq:marglik}}\label{app:marglik}
From Bayes' rule, we have
\begin{align}
p(x_{1:T}\mid \xi) = \frac{p(A,Q\mid\xi)p(x_{1:T}\mid A,Q,\xi)}{p(A,Q\mid \xi, x_{1:T})}.
\end{align}
The expression for each term is found in~(\ref{eq:MNpdf}-\ref{eq:MNIW}), \eqref{eq:xlik} and~\eqref{eq:mnpost}, respectively. All of them have a functional form $\eta(\xi)\cdot|Q|^{\chi(\xi)}\cdot\exp\left(-\frac{1}{2}\tra{Q^{-1}\tau(A,x_{1:T},\xi)}\right)$, with different $\eta, \chi$ and $\tau$. Starting with the $|Q|$-part, the sum of the exponents for all such terms in both the numerator and the denominator sums to $0$. The same thing happens to the $\exp$-part, which can either be worked out algebraically, or realized since $p(x_{1:T}\mid \xi)$ is independent of $Q$. What remains is everything stemming from $\eta$, which indeed is $p(x_{1:T}\mid \xi)$,~\eqref{eq:marglik}.

\subsection{Invariant distribution of Algorithm~\ref{alg:gibbs}}\label{app:mh-w-g}
As pointed out by \citet{DJ:2014}, the combination of Metropolis-within-Gibbs and partially collapsed Gibbs might obstruct the invariant distribution of a sampler. In short, the reason is that a Metropolis-Hastings (MH) step is conditioned on the previous sample, and the combination with a partially collapsed Gibbs sampler can therefore be problematic, which becomes clear if we write the MH procedure as the operator $\mathcal{MH}$ in the following simple example from \citet{DJ:2014} of a sampler for finding the distribution $p(a,b)$:
\begin{algorithmic}\small
	\State
	Sample $a[k\!+\!1]	\sim p(a \mid b[k])$ (Gibbs)
	\State
	Sample $\mathrlap{b}\phantom{a}[k\!+\!1] \sim \mathcal{MH}(b \mid a[k\!+\!1], b[k])$ (MH)
\end{algorithmic}
So far, this is a valid sampler. However, if collapsing over $b$, the sampler becomes
\begin{algorithmic}\small
	\State
	Sample $a[k\!+\!1]	\sim p(a)$ (Partially collapsed Gibbs)
	\State
	Sample $\mathrlap{b}\phantom{a}[k\!+\!1] \sim \mathcal{MH}(b \mid a[k\!+\!1], \nb{b[k]})$ (MH)
\end{algorithmic}
where the problematic issue, obstructing the invariant distribution, is the joint conditioning on $a[k\!+\!1]$ \textit{and} $b[k]$ (marked in red), since $a[k\!+\!1]$ has been sampled without conditioning on $b[k]$. Spelling out the details from Algorithm~\ref{alg:gibbs} in Algorithm~\ref{alg:gibbs2}, it is clear this problematic conditioning is not present.
\begin{algorithm}
	\caption{Details of Algorithm~\ref{alg:gibbs}}
	\begin{algorithmic}[1]\small
		\makeatletter
		\setcounter{ALG@line}{1}
		\makeatother
		\For{$k = 0$ to $K$}
		\State
		Sample ${{x}_{1:T}[k\!+\!1]}
		\boldsymbol{\:\big\vert\:}
		\mathrlap{{A}[k],{Q}[k],\xi[k] }$\phantom{${x}_{1:T}[k\!+\!1], \xi[k\!+\!1], {Q}[k\!+\!1]$} (Gibbs)
		\State
		Sample  \phantom{${x}_{1:T}[k\!+\!1]$}$\mathllap{{\xi}[k\!+\!1]} \sim \mathcal{MH}(x_{1:T}[k\!+\!1],\xi[k])$
		\State
		Sample \phantom{${x}_{1:T}[k\!+\!1]$}$\mathllap{{Q}[k\!+\!1]}
		\boldsymbol{\:\big\vert\:}
		\mathrlap{\xi[k\!+\!1],{x}_{1:T}[k\!+\!1]}$\phantom{${x}_{1:T}[k\!+\!1], \xi[k\!+\!1], {Q}[k\!+\!1]$}(Gibbs)
		\State
		Sample \phantom{${x}_{1:T}[k\!+\!1]$}$\mathllap{{A}[k\!+\!1]}
		\boldsymbol{\:\big\vert\:}
		\mathrlap{{Q}[k\!+\!1], \xi[k\!+\!1], {x}_{1:T}[k\!+\!1]}$\phantom{${x}_{1:T}[k\!+\!1], \xi[k\!+\!1], {Q}[k\!+\!1]$}(Gibbs)
		\EndFor
	\end{algorithmic}
	\label{alg:gibbs2}
\end{algorithm}

\newpage
\small
\bibliographystyle{abbrvnat}       
\bibliography{../../references}           

\end{document}